# Employee Performance when Implementing Agile Practices in an IT Workforce

MUHAMMAD HAMID RAZA, MHR, MOOKADAM

Department of Informatics, University of Pretoria, Pretoria, Gauteng, South Africa, hr.mookadam@gmail.com

RIDEWAAN, R, HANSLO

Academy of Computer Science and Software Engineering, University of Johannesburg, Johannesburg, Gauteng, South Africa, ridewaanh@uj.ac.za

Adoption of agile practices has increased in IT workforces. However, there is a lack of comprehensive studies in the African context on employee performance when implementing agile practices. This study addresses this gap by exploring employee performance in agile environments for IT workforces in South Africa. An interpretivist mono-method qualitative approach was used, with the use of interviews as a research strategy. Seventeen semi-structured interviews were conducted with agile practitioners from various roles. Our results indicated that agile practices influence employee performance significantly, with participants reporting on aspects which included planning, communication, employee development and well-being, collaboration, team culture and progress. Additionally, our results reported obstacles when using agile practices that included adoption, team engagement, leadership and instilling an agile mindset. Agile practices influence employee performance in IT workforces by fostering improved team dynamics, enhanced collaboration, improved efficiencies, risk management, planning, continuous improvement, learning, personal development and well-being. Conclusively, our findings suggest that if agile challenges are addressed and additional support is provided, employee performance can be significantly improved.

**CCS CONCEPTS** • Software and its engineering • Software development process management • Agile software development

**Additional Keywords and Phrases:** Agile adoption, Agile mindset, Employee well-being, Expectancy Theory, Agile coaching

## 1 INTRODUCTION

Agile practices consist of a specific outline of principles and values involved in the software development life cycle (SDLC), which enables software development teams to create value for the business by delivering working software in smaller pieces. The implementation of agile in the workplace, within the software development phase, creates an environment which encourages change and adaptation [1]. Agile practices follow four core values: the value of "Individuals and interactions over processes and tools"; "working software over comprehensive documentation"; "customer collaboration over contract negotiation" and "responding to change over following a plan" [2].

Due to the IT industry experiencing substantial project failures that were caused by inefficiencies in the SDLC, agile was founded in February 2001. According to the Standish Group's 2020 CHAOS Report, approximately 66% of software development projects were unsuccessful. Increased employee performance in the workforce is considered crucial to the success of the organization [3]. Hence, the methodology implemented in the workforce must create an environment contributing to this.

Limited research has been done in determining employee performance when agile is used, and more specifically, within an African IT workforce. A major concern is that the implementation of agile could either motivate employees to work better or hinder their performance within the organization. To make an informed decision about implementation, South African IT organizations must assess the impact of agile practices on employee performance. The objective is to ensure that such methodologies are mutually beneficial for the software development process and the employees, who are the primary agents of success.

The primary aim of this research is to analyze the relationship between the adoption of agile practices and employee performance within the South African IT sector. By providing practical recommendations, the study intends to guide organizations in optimizing performance during agile transformations. The investigation is therefore guided by the following research question: "*How does the implementation of agile practices in a South African IT workforce contribute to employee performance?*"

This paper contains six sections. The literature review (section 2) follows this introduction. Section 3 unpacks the research methodology, and Section 4 provides the research findings. Section 5 discusses the findings, and Section 6 concludes this paper.

## 2 LITERATURE REVIEW

### 2.1 Key characteristics of agile

There are multiple characteristics of agile. Some of these key agile characteristics are explained below:

- **Iterative**. Agile focuses on increasing customer value. This fosters agile teams to focus on a portion as opposed to the entire project, in cycles which have a defined goal [4].
- **People-oriented**. The agile value of people over processes is key to the individuals involved in system development processes. This prioritizes people over tools, which puts the end-user as the focus and allows the team to evolve and potentially increase employee performance and productivity [4].
- **Adaptive**. Agile processes are built on an iterative method, and many unknowns or risks may occur. The agile approach prepares the team for these scenarios and how to react [4].
- **Collaborative approach**. With agile, the development approach is based on different iterations and developing different modules that incrementally lead to the final software. This requires collaboration and communication from the team to ensure that progress moves flawlessly [5].

### 2.2 Benefits and Challenges of Agile Practices

Agile practices were founded in 2001, and as the years went by, software development teams could understand the pros and cons that agile brings compared to traditional methodologies. Dima and Maasen [6] conducted a study where 19 professionals who worked in the software development field based at IT companies globally were interviewed. During the interview process, when asked about the advantages and disadvantages that using agile brings to the software development team, 77% and 85% of the respondents mentioned frequent useful feedback and better adaptability to meeting requirements, respectively, as advantages of agile. With regard to the disadvantages being noted by the respondents, 76% mentioned that extra pressure on the team is created.

*2.2.1 Benefits*

Some of the benefits of agile are identified below:
- **Increased employee competition**. With agile being a fast-paced delivery, this creates an environment of competitiveness amongst team members with regard to completing high-quality tasks on time [6]
- **Improved team cooperation**. The ability for employees to work within groups and with diverse teams that have the same goal in mind [6]
- **Ensure customer satisfaction**. Agile requires close end-user and customer engagement, and this allows the software system to be evaluated by them and provides feedback to the development team to improve the system after each iteration [5]
- **Quick releases**. This allows the software development team to release working software earlier in the process to provide value to the business earlier in the project [7].

*2.2.2 Challenges*

Like all software development methodologies, agile also has its challenges. Some of these challenges are described below:
- **Extra pressure on employees**. With the high pace that agile brings to the team with the agile software development process, software release and improvements communicated by end-users, this constant cycle increases the pressure within the team [6].
- **Lack of documentation**. With the Waterfall methodology that focuses on intense documentation that is distributed to the development team, agile is low on documentation, and if the project tends to be complex, the lack of documentation can be a disadvantage [8].
- **Staff turnover rate due to high work pressure**. The increased pressure that agile has on its employees tends to be a factor in the staff turnover rate [6].

## 2.3 Factors Influencing Employee Performance

There have been many studies done on employee performance in organizations in multiple industries. However, there have been limited studies on employee performance in an agile environment, and the IT workforce in Africa specifically. Diamantidis and Chatzoglou [9] break down employee performance into three elements: firm/environment-related factors, job-related factors and employee-related factors.
- **Firm/environment-related factors**. Factors within this grouping are directly linked to an organization and the way that it operates. These factors include an organization's stance on employee training, leadership and management and the impact these have on employee performance [9][10].
- **Job-related factors**. These factors relate to the job or task that an employee carries out. This includes the environment the employee works in, the communication between their colleagues and management and the control and autonomy that is given to an employee [9][11].
- **Employee-related factors**. These are factors on the employees themselves that are instilled within them that have an impact on how they perform their duties, which ultimately has an impact on their performance. These factors include their skill level, attitude, and whether they can complete duties separate from their current roles [9][12].

## 2.4 Acceptance of Agile Practices by Employees

The implementation of a methodology comes with constraints in that there is resistance from the team and organizations to accepting and adopting a methodology. Past research demonstrates that when organizations change from traditional methodologies to agile, it has a huge impact on organizations and their employees. It does not just require changing existing tools and technologies; it impacts multiple components such as organization structure, management behavior, the team and organizational culture shift and having the buy-in from all stakeholders [13]. For employees to accept agile, an organization needs to provide its employees with the necessary training and support and instill an organizational culture that changes employee behavior and allows for knowledge management to transpire. Key aspects that lead to acceptance, as defined by Chan and Thong [22], are ability-related factors, which refer to the team members' abilities themselves and the support and coaching provided to them to use agile practices, as well as motivation-related factors, which relate to organizational culture and high management support in promoting the use of agile practices to its employees, and opportunity-related factors, which refer to the opportunity for employees to create and transfer knowledge.

## 2.5 Theoretical Framework for Employee Performance

Vroom's Expectancy theory is used as a theoretical lens in this study. Vroom's Expectancy is applied to employee motivation, which indicates a correlation between behavior and the expected outcome of that behavior which individuals believe to have a positive outcome or reward [14]. Three characteristics in which individuals evaluate their behavior are:

**Valence**. This is the value an individual place on the outcome. For an employee to be motivated, the reward must align with what they value so that their increased effort will improve their performance. Within an agile context, employees may value the ability to be flexible and innovative, or they may value the decrease in stress and enhanced collaboration among team members. Rewards ultimately differ from person to person.

**Instrumentality**. This refers to the belief that an individual will be rewarded for their increased performance if they meet the expectation of a task. To improve performance, managers should ensure that employee performance and rewards are clearly aligned. This creates an understanding for an employee that high performance will lead to rewards, which could be both intrinsic and extrinsic.

**Expectancy**. This is the assumption that performing a behavior will result in a successful outcome. It means an individual believes that the effort they put in can affect the performance they deliver. For example, if an employee believes they put in the effort to understand and apply agile practices, this will lead to increased levels in their performance. This can be negatively affected by a lack of organizational support, such as inadequate training on agile practices.

The strength of the Expectancy theory has been demonstrated by other studies to predict employee performance [15][16].

## 3 METHODOLOGY

### 3.1 Research Design

The chosen paradigm for this study was interpretivism. The reason is that the study aimed to explore the influence employee performance has when the agile practices are implemented in South African IT workforces. Interpretive research often uses qualitative methods, and this study uses interviews as a research strategy that allows the researchers to obtain data from different individuals in different social phenomena [17]. In addition, this research study used the inductive research approach, which is commonly used within qualitative methods. This assisted the researchers in gaining an interpretation of the perceptions of individuals working in an agile environment and their behavior towards employee performance.

### 3.2 Data Collection

The primary data collection method for this study was semi-structured interviews, also referred to as qualitative research interviews. This allowed the authors to understand what was happening in a specific reality to seek new knowledge. Although semi-structured interviews involved the researcher having pre-defined open-ended questions, this also provided the authors with the freedom to probe participants into exploring topics that emerged during the interview [18]. This study focuses on taking an interest in a phenomenon at a particular time, thus, a cross-sectional design is used.

### 3.3 Sampling

The target participants for this study were agile practitioners who worked within IT workforces in South Africa, and more specifically, the software development field. For the purpose of this study, a multi-sampling technique was used, which included purposive and snowball sampling techniques. The purposive sampling technique allowed the researchers to target participants who have experience working in agile software development environments in South African IT workforces. Furthermore, the snowball sampling technique was also used to help identify additional participants. Saturation in interviews generally occurs after the twelfth interview, but it is imperative to do a few more interviews thereafter to verify saturation [19][20]. The initial estimate for this study was 15, however, we continued to 17 until no further data was discovered.

### 3.4 Data Analysis

The data analysis method selected for this study was thematic analysis, which is often used in qualitative methods. It allowed us to explore and describe the data by identifying, analyzing and reporting on patterns and themes from the interview data. This study was guided by six phases that are outlined by Braun and Clarke [20] as (1) familiarizing yourself with the data, (2) generating initial codes, (3) searching for themes, (4) reviewing themes, (5) defining and naming themes, and (6) producing the report.

### 3.5 Ethical Considerations

All participants involved should be treated fairly and with honesty. A consent form was provided to the potential participant beforehand and indicated the purpose and description of the study, how the data would be used, any risks and benefits to the participant, confidentiality of the participant and the data collected, and the storage process [21]. For this study, no personal information of participants was requested.

# 4 FINDINGS

## 4.1 Description of Participants

The interviews conducted included participants who were employed at various South African IT organizations. These participants worked in software development environments, where agile practices were used. To ensure data triangulation, participants with different role types with different seniority levels were used (Table 1).

Table 1: Demographics of participants

| Research participant: Agile practitioners | Role type | Years of experience in agile |
|---|---|---|
| P1 | Business analyst | 8 |
| P2 | Test analyst | 6 |
| P3 | Software developer | 3 |
| P4 | Software developer | 6 |
| P5 | Business architect | 7 |
| P6 | Software development manager | 12 |
| P7 | Scrum master | 8 |
| P8 | Business analyst | 3 |
| P9 | Business architect | 8 |
| P10 | Software development manager | 10 |
| P11 | Software developer | 9 |
| P12 | Systems architect | 6 |
| P13 | Business architect | 5 |
| P14 | Business analyst | 3 |
| P15 | Product owner | 10 |
| P16 | Project manager | 5 |
| P17 | Chief information officer | 11 |

## 4.2 Emerging Themes

Through the analysis of participants' responses from the interviews, a total of seven themes emerged (Figure 1).

### 4.2.1 Theme 1: Agile Practice Benefits.

There are various benefits of agile practices that were identified. Nine secondary themes emerged: (1) team dynamics and collaboration, (2) communication, (3) efficiency and productivity, (4) quality and delivery, (5) risk management and problem solving and (6) planning and adaptability, (7) continuous improvement and learning, (8) project management and (9) personal development and well-being. The most frequently mentioned benefit was team collaboration.

### 4.2.2 Theme 2: Agile Practice Challenges.

There were numerous challenges of agile practices that were identified. Five emerging secondary themes from the primary theme: (1) adopting agile practices, (2) leadership and decision making, (3) team engagement, (4) customer involvement and (5) planning and predictability. The most cited challenges were related to adopting agile practices and fostering an agile mindset.

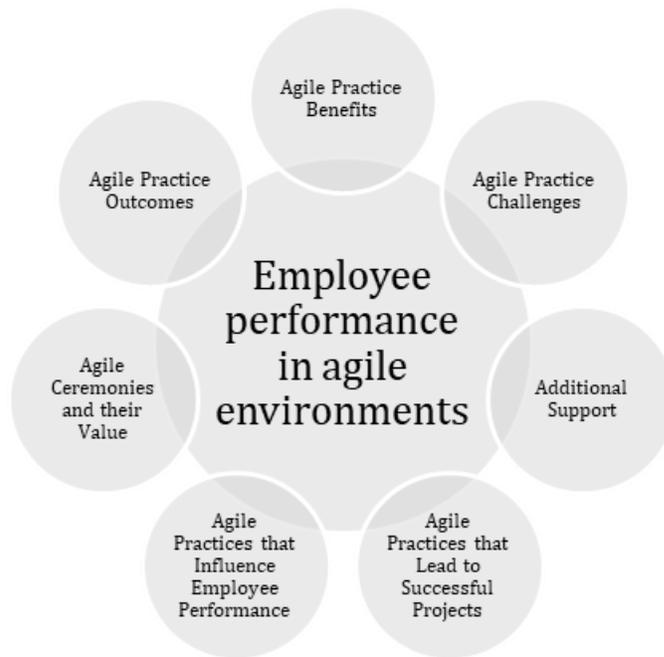

Figure 1 Employee performance in agile environments: emerging themes

*4.2.3 Theme 3: Additional Support.*

Various additional support requirements for using agile practices more effectively were identified. Four secondary themes were identified: (1) training and education, (2) additional role support, (3) acceptance and understanding and (4) implementation and practices. The need for an agile coach was a prominent support requirement mentioned by participants.

*4.2.4 Theme 4: Agile Practices that Lead to Successful Projects.*

Analysis of the results described the different aspects of agile practices that led to successful projects. Five secondary themes emerged: (1) team dynamics and collaboration, (2) communication, (3) agile practices and processes, (4) planning and management and (5) problem solving. Planning and preparation, along with defining a minimum viable product (MVP), were key aspects.

*4.2.5 Theme 5: Agile Practices that Influence Employee Performance.*

Features of agile practices were identified that influenced employee performance. Five secondary themes emerged: (1) task management and planning, (2) transparency and communication, (3) employee development and well-being, (4) team collaboration and culture and (5) progress and outcomes. Planning was the most frequently mentioned aspect influencing performance.

*4.2.6 Theme 6: Agile Ceremonies and their Value*

The following agile ceremonies were found to be used by participants: (1) daily stand-up, (2) sprint planning, (3) backlog/sprint grooming, (4) sprint review, (5) sprint retrospective, and (6) knowledge and sharing. All participants used daily stand-ups and sprint planning. Participants reported that these ceremonies provided value in areas such as tracking progress, identifying issues, improving communication, planning and estimation, and fostering continuous improvement.

*4.2.7 Theme 7: Agile Practice Outcomes.*

Participants provided feedback relating to the existing and motivational outcomes obtained when using agile practices. For existing outcomes, five key aspects were identified: recognition, team dynamics, sense of achievement and accomplishment, professional development, product delivery and business value. For motivational outcomes, six key aspects were identified: personal growth and development, work satisfaction and accomplishment, respect and collaboration, agile practices and mindset, flexibility and work-life balance, and recognition and value addition.

Multiple forms of valued and undervalued outcomes were identified. Four key aspects of valued outcomes were identified: recognition and feedback, team collaboration and culture, personal and professional growth, and delivery and business impact. Three key aspects of undervalued outcomes were identified: recognition, team dynamics, and feedback and individual value. A key undervalued outcome was the appreciation of work within the team.

## 5 DISCUSSION OF FINDINGS

### 5.1 Discussion of Agile Benefits and Challenges

Agile practices offer various benefits that have a positive influence on employee performance, and agile also presents challenges that hinder performance. Additional support mechanisms can help teams navigate the complexities of using agile practices to enhance employee performance.

Our results found that agile practices have substantially enhanced team dynamics and collaboration. Participants reported that agile practices provide a setting that increases a sense of unity amongst the team. As P7 stated, "*the way we work has changed from being in silos to working and being more collaborative within the team*." This is attributed to working towards a common goal and delivering as a team. Results from participants found that communication is a key component in agile practices, with P2 noting, "*A benefit of agile is communication that was quite important. Agile encourages more communication between different members of the team and it builds a good relationship with one another within your team*." Agile practices are also set up to increase efficiency, which assists teams in increasing productivity. P3 mentioned, "*you have a clear outline of what you're going to do when the sprint begins and what you are working towards and this increases productivity*."

However, agile adoption raises a few challenges. One participant (P12) highlighted the difficulty of adoption, stating, "*Adoption takes time when people are conditioned in the way that they work to bring a new methodology... some individuals are resilient, and others are resistant to change*." Effective leadership support is vital, as poor leadership can be a significant obstacle. P6 bluntly stated, "*I've seen many transformations fail because of dictatorship*."

Providing agile teams with adequate training and educational resources is fundamental for successful adoption. For agile teams to implement agile practices effectively, it requires guidance and mentorship. Our results show that an involved agile coach is highly beneficial, as P13 mentioned, it is important "to have an agile coach initially that will guide you on how things need to be done."

### 5.2 Implementation of Agile Practices through Agile Ceremonies

Agile practices are designed to facilitate the software development life cycle by conducting various agile ceremonies that guide team dynamics, collaboration, communication, and planning. Our results show the various agile ceremonies that participants use, including daily stand-up meetings, sprint planning, sprint retrospectives, and backlog grooming.

Daily stand-ups provide significant value to teams to understand progress and identify blockers. As P9 explained, "*It is beneficial during stand-up to understand if there are any blockers or any issues that need to be resolved as soon as possible so that the team can continue working*." Sprint planning sessions are important to set the stage for a successful sprint. A principle of agile practices is for teams to reflect regularly, and sprint retrospectives allow them to do this. P17 noted that a retrospective "*is an opportunity for the team to reflect back on how they can improve*."

### 5.3 Contribution of Agile Practices to Employee Performance

Agile practices put a high focus on the human element during software development. Agile teams' ability to conduct effective planning and manage tasks efficiently are key features that contribute to employee performance. As P2 noted, with agile, "*you are able to manage your time better and because of that you able to actually produce better*."

A principle of agile practices is to build agile teams with motivated individuals who are supported and given the level of autonomy necessary to get work done. P9 highlighted this, saying, "*In an agile environment, teams have more autonomy and more say in the type of features that are going into a sprint. They have a lot more influence and that can improve their performance*." This culture is essential to promote personal growth and well-being.

Agile practices are also designed to frequently track progress and outcomes. This frequent delivery of tangible software displays a sense of progress to the team that builds momentum. As P11 put it, "*you get that sense of progress as opposed to the waterfall process sitting on a project for three 4-5 months and then only seeing the fruits of your labour*."

### 5.4 Theoretical Framework Alignment

This study used the Expectancy Theory as a lens to help answer the research questions. The theory is key to understanding employee performance, suggesting a correlation between an individual's behavior and the expected outcome of that behavior [14]. It is grouped into three variables: expectancy, instrumentality, and valence [15].

- Concerning the **expectancy** variable, it is the belief that increased effort will lead to improved performance. Agile practices promote increased effort through benefits like improved team dynamics and continuous improvement.
- **Instrumentality** is the belief that improved employee performance will be rewarded with an outcome. Our results show that recognition from peers and customers, a sense of achievement, and professional development are key outcomes participants take away from using agile practices.

- **Valence** is the value that an individual places on the outcome. Our results show that recognition, an enhanced team culture, personal growth, and delivering successful projects are highly valued outcomes. As P6 mentioned, a positive team environment creates "*a good mental shift around the team*."

## 6 CONCLUSION

The purpose of this study was to explore agile practices and their contribution to employee performance in South African IT workforces. The findings of agile benefits when using agile practices were team dynamics and collaboration, communication, efficiency and productivity, quality and delivery, risk management and problem-solving, planning and adaptability, continuous improvement and learning, project management, and personal development and well-being. The findings of agile challenges when using agile practices were the adoption of agile practices, leadership and decision making, team management, customer involvement, and planning and predictability. During the analysis, the influence agile practices have on employee performance was task management and planning, transparency and communication, employee development and well-being, team collaboration and culture, and progress and outcomes.

This study contributes to academia on agile practices by providing evidence on how employee performance is influenced by agile practices within South African IT workforces. From a practical perspective, this study contributes to South African IT organizations by providing insight into how effectively implementing agile practices enhances employee performance.

It is important to note the limitations experienced during this study. This study was restricted to employees using agile practices within South African IT workforces. Therefore, the findings cannot be generalized and may not apply to IT workforces outside of South Africa. The sample size was also limited to seventeen participants, which might not represent the broader IT workforce accurately.

For future research, researchers can create a conceptual framework using the Expectancy Theory as a basis for understanding how agile practices influence employee performance. By providing the relevant working conditions, additional support, addressing challenges, and using agile practices effectively, it can increase employees' efforts, which will result in increased performance. Increased employee performance can be maintained if outcomes obtained align with the outcomes that employees value.